\documentclass[showpacs,twocolumn,preprintnumbers,amsmath,amssymb]{revtex4}
\usepackage{amsmath,amsfonts,latexsym,amssymb,graphicx,graphics,epsfig,subfigure,color,makeidx}
\usepackage{xcolor,diagbox}
\usepackage{multirow}
\usepackage[colorlinks,linkcolor=blue,anchorcolor=blue,citecolor=green,urlcolor=blue]{hyperref}
\usepackage{mathrsfs}
\usepackage{amsmath}

\begin{document}
\title{Dark Information in Black Hole with $\lambda\varphi$ Fluid}

\author{Yu-Xiao Liu$^{a}$\footnote{liuyx@lzu.edu.cn}}
\author{Yu-Han Ma$^{b}$$^{c}$\footnote{yhma@@gscaep.ac.cn}}
\author{Yong-Qiang Wang$^{a}$\footnote{yqwang@lzu.edu.cn}}
\author{ Shao-Wen Wei$^{a}$ \footnote{weishw@lzu.edu.cn}}
\author{Chang-Pu Sun$^{b}$$^{c}$\footnote{suncp@gscaep.ac.cn}}

\affiliation{$^{a}$Research Center of Gravitation $\&$ Lanzhou Center for Theoretical Physics,
School of Physical Science and Technology, Lanzhou University, Lanzhou 730000, China\\
$^{b}$ Graduate School of China Academy of Engineering Physics,
     No. 10 Xibeiwang East Road, Haidian District, Beijing  100193, China\\
$^{c}$Beijing Computational Science Research Center, Beijing 100193, China}

\begin{abstract}
It has been shown that the nonthermal spectrum of Hawking radiation will lead to information-carrying correlations between emitted particles in the radiation. The mutual information carried by such correlations can not be locally observed and hence is dark. With dark information, the black hole information is conserved. In this paper, we look for the spherically symmetric black hole solution in a $\lambda\varphi$ fluid model and investigate the radiation spectrum and dark information of the black hole. The spacetime structure of this black hole is similar to that of the Schwarzschild one, while its horizon radius is decreased by the $\lambda\varphi$ fluid. By using the statistical mechanical method, the nonthermal radiation spectrum is calculated. This radiation spectrum is very different from the Schwarzschild case at its last stage because of the effect of the $\lambda\varphi$ fluid. The $\lambda\varphi$ fluid reduces the lifetime of the black hole, but increases the dark information of the Hawking radiation.
\end{abstract}

\maketitle

\section{Introduction}

{The presence of black holes is one of the important predictions of general relativity, and this prediction has been proven to be true by the LIGO-Virgo collaborations through the direct observation of gravitational waves \cite{LIGOScientific:2016aoc}. In recent years, there are a large number of works on the study of black holes, such as the black hole shadow, deflection angle, quasinormal modes, thermodynamic phase transitions and other topics
\cite{Liu:2020yqa,Cai:2021sag,Wei:2021bwy,Wang:2021art,Cimdiker:2021cpz,Okyay:2021nnh,Ovgun:2021ttv,Long:2020wqj,Cadoni:2021jer,Javed:2019ynm}.}

It is well known that gravitation, quantum theory, and thermodynamics {are connected} deeply by Hawking radiation of black holes and a lot of success has been achieved. However, black hole radiation also raises some puzzles. One serious puzzle is the information loss paradox proposed by Hawking \cite{Hawking1976paradox}. By using the semiclassical approximation, he found that the emitted radiation is exactly thermal and is determined only by the geometry of the black hole outside the horizon. Therefore, the radiation has nothing to do with the detailed structure of the body that {collapses} to form the black hole. Since there are correlations between the accessible degrees of freedom outside the horizon and the inaccessible degrees of freedom behind the horizon, the radiation detected by observers outside the horizon is in a mixed state. After the black hole completely evaporates, the radiation is the whole system. Therefore, an initially pure quantum state of the body that can be precisely known has evolved to a mixed state that cannot be predicted with certainty. {However, this contradicts with the unitarity of operators required by quantum mechanics, for which the evolution of a pure state to a mixed state is forbidden.} This is the information loss paradox \cite{Preskill9209058}.

In order to solve the information loss puzzle, some resolutions have been proposed.
In some resolutions, the information could come out with the Hawking radiation and all of the information could come out at the end of the Hawking radiation. In other resolutions, it could be retained by a stable black hole remnant \cite{Giddings1992} or be encoded in ``quantum \mbox{hair'' \cite{Horowitz1988,Wilczek1989,Wilczek1991}.} The information even can escape to a ``baby universe''~\cite{Zeldovich1977,Hawking1990}. However, there was no satisfactory resolution at an early stage \cite{Preskill9209058}. Later development indicates that this paradox can be resolved in string theory by a new picture of black holes: fuzzballs, {which describe black hole microstates \cite{Mathur2002}.} In this fuzzball paradigm, the black hole is replaced by an object {without a horizon} and singularity. Besides, there are other ideas such as firewalls \cite{Polchinski2013}, {entanglement \cite{Braunstein2009}, island and entanglement wedge \mbox{reconstruction \cite{Penington:2019npb,Almheiri:2019yqk,Almheiri:2019hni}}.}

In 2000, Parikh and Wilczek \cite{Wilczek2000} presented a consistent derivation of Hawking radiation as a tunneling process and found that the Hawking radiation spectrum is nonthermal because of conservation laws. This nonthermality of the radiation allows the possibility of information-carrying correlations between subsequently emitted particles in the radiation.

Zhang and Cai et al. \cite{ZhangCai2009} discovered correlations among Hawking radiations from a black hole by using standard statistical method. Then, by considering the mutual information carried by such correlations, they found that the black hole evaporation process is unitary and the black hole information is conserved. It was found that there is an even deep origin of nonthermal nature of Hawing radiation without referring to the horizon geometry \cite{Dong2014}.
Recently, by considering the canonical typicality, Ma and Sun et al. \cite{MaChenSun:2018EPL} showed that the nonthermal radiation spectrum is independent of the detailed quantum tunneling dynamics, and the black hole information paradox could be naturally resolved with the correlations between the black hole and its radiation.

The correlation information is dark because it can not be locally observed in principle even though the Hawking radiation can be finally measured experimentally \cite{MaChenSun:2018NPB}. This information is called dark information, which can be measured nonlocally only with two or more detectors. Such coincidence measurement is similar to the Hanbury--Brown--Twiss experiment for the coincidence counting in quantum optics \cite{Scully1999}.

Recently, the influence of dark energy on black hole radiation and dark information was studied in~\cite{MaChenSun:2018NPB} by the approach of canonical typicality. It was found that, with the existence of  dark energy, the black hole has lower Hawking temperature and hence longer lifetime. Furthermore, dark energy will enhance the nonthermal effect of the black hole radiation and raise the dark information of the radiation~\cite{MaChenSun:2018NPB}.

{It is well known that dark matter and dark energy compose about $27\%$ and $68\%$ of our universe, respectively. Dark matter affects and accounts for the evolution of our universe, the formation of large-scale structure, and galaxy rotation curves \cite{Vagnozzi2017}. Especially, there is a great deal of dark matter in each galaxy. As the case of dark energy, we have also many candidates for dark matter, such as weakly interacting massive particles (WIMPs), super WIMPs, light gravitinos, sterile neutrinos, hidden dark matter, axions, and primordial black holes \cite{Feng2010,Carr2016,Huang2018,Bartolo2019}.
In this paper, inspired by the work~\cite{MaChenSun:2018NPB}, we would like to investigate the effect of the $\lambda\varphi$ fluid on the Hawking radiation and dark information of the black hole.
The $\lambda\varphi$ fluid theory was proposed in~\cite{Lim:2010yk} to give a unified description of dark matter and dark energy. This field theory contains two scalar fields, but has only one single degree of freedom. In this theory, the fluid velocity is always tangent to geodesics and hence it can mimic ``dust". However, unlike a standard cold dark matter fluid, the $\lambda\varphi$ fluid carries pressure parallel to its fluid velocity. We will consider such $\lambda\varphi$ fluid and find the solution of an uncharged spherically symmetric black hole. Then, we will study the Hawking radiation and dark information for the black hole with the $\lambda\varphi$ fluid.}

This paper is organized as follows. In Section~\ref{Setup}, we review the {the $\lambda\varphi$ fluid theory} briefly and derive the Einstein equations for a spherically symmetric metric. In Section~\ref{solution}, we look for an analytical black hole solution in an asymptotic flat spacetime and analyze its properties. In Section~\ref{TemperatureEntropy}, we calculate the black hole mass, temperature and entropy. Then, the radiation spectrum and dark information of the black hole with {the $\lambda\varphi$ fluid} are calculated in Section~\ref{RadiationSpectrum} and Section~\ref{DarkInfor}, respectively. In the end, conclusions and discussions are given in Section~\ref{conclusion}.

\section{Einstein Equations in {the $\lambda\varphi$ Fluid Theory}}\label{Setup}

{
In this section, we will give a brief introduction to the $\lambda\varphi$ fluid theory and derive the Einstein equations for a spherically symmetric spacetime.
We consider the following action
\begin{eqnarray}
S={M_{\text{Pl}}^2} \int   d^4x\sqrt{-g}\left[ \frac{1}{2}R + L(\lambda,\varphi,X)\right], \label{actionfluid}
\end{eqnarray}
where the Lagrangian $L(\lambda,\varphi,X)$ for the scalar fields is described by the $\lambda\varphi$ fluid theory~\cite{Lim:2010yk}:
\begin{eqnarray}
L(\lambda,\varphi,X)=K(\varphi,X) + \lambda \left(-2X-U(\varphi)\right) \label{Lagrangian1}.
\end{eqnarray}

Here, $X$ is a standard kinetic term for the scalar field $\varphi$, $X=-\frac{1}{2}\partial^{\mu}\varphi\partial_{\mu}\varphi$, $\lambda$ is a ``Lagrange multiplier'' having no kinetic term, $K$ is a function of $\varphi$ and $X$, and $U(\varphi)$ is a function of $\varphi$. It can be seen that the perssure is identically vanishing without the term $K$.
The Lagrange multiplier $\lambda$ enforces a constraint between the value of the scalar field $\varphi$ and the norm of its derivative. Thus, the dynamics of the $\lambda\varphi$ fluid is determined by two first-order ordinary differential equations and there are no propagating wave-like degrees of freedom \cite{Lim:2010yk}.
}

{It is worth noting that the mimetic gravity proposed in \cite{Chamseddine:2013kea} is in fact a special case of the above $\lambda\varphi$ fluid theory. The action for the mimetic gravity is given by
\begin{eqnarray}
S\!=\!{M_{\text{Pl}}^2}\!\int \!\! d^4x\sqrt{-g}\left[ \frac{1}{2}R\!+\!\lambda
      \left(\partial^{\mu}\varphi\partial_{\mu}\varphi+1\right)\right],
\end{eqnarray}
which can be obtained from (\ref{actionfluid}) and (\ref{Lagrangian1}) by taking $K=0$ and $U=1$. In this theory, the conformal degree of freedom of the metric is isolated in a covariant way.
This is done by rewriting the physical metric $g_{\mu\nu}$ in terms of an auxiliary metric $\tilde g_{\mu\nu}$ and a scalar field $\varphi$~\cite{Chamseddine:2013kea}. The explicit relation between them is given by
\begin{eqnarray}\label{relationofpanda}
g_{\mu\nu}=-\tilde g_{\mu\nu}\tilde g^{\alpha\beta}\partial_{\alpha}\varphi\partial_{\beta}\varphi.
\end{eqnarray}

As a consequence, the scalar field satisfies the following constraint
\begin{eqnarray}\label{conditionofphi}
g^{\mu\nu}\partial_{\mu}\varphi\partial_{\nu}\varphi=-1.
\end{eqnarray}

It is shown that, under the conformal transformation of the auxiliary metric: $\tilde g_{\mu\nu}\rightarrow \Omega^2(x^{\alpha})\tilde g_{\mu\nu}$ with $\Omega(x^{\alpha})$ a function of the spacetime coordinates, the physical metric is invariant. For a review  {on the mimetic gravity, see~\cite{1612.08661}.}
}

In this paper, we consider the simple case of $K=-V(\varphi)$ since the kinetic term $X$ is constained to be $X=-U(\varphi)/2$ by the Lagrange multiplier. Thus, the Lagrangian (\ref{Lagrangian1}) can be rewritten as
\begin{eqnarray}
L(\lambda,\varphi,X)= \lambda \left( \partial^{\mu}\varphi \partial_{\mu}\varphi-U(\varphi)\right) -V(\varphi). \label{Lagrangian2}
\end{eqnarray}

The equations of motion (EoMs) are obtained by varying the above action (\ref{actionfluid}) with respect to $g_{\mu\nu}$, $\varphi$ and $\lambda$, respectively:
    \begin{eqnarray}
        \label{var eom1}
        G_{\mu\nu}+2\lambda \partial_\mu \varphi \partial_\nu \varphi-L g_{\mu\nu}=0,    \\
        \label{var eom2}
        2\lambda\Box\varphi+2\nabla_{\mu}\lambda~\nabla^{\mu}\varphi+\lambda U_{\varphi}+V_{\varphi}=0, \\
        \label{var eom3}
        g^{\mu\nu}\partial_\mu \varphi \partial_\nu \varphi-U=0.
    \end{eqnarray}

Here the notation $P_{\varphi}$ is defined as $P_{\varphi} \equiv \partial P/\partial\varphi$ and the d'Alembert operator is given by $\Box \equiv g^{\mu\nu}\nabla_{\mu}\nabla_{\nu}$.

We consider the following spherically symmetric metric
    \begin{eqnarray}
        \label{metric1}
       ds^2=-k(r)dt^2+ \frac{dr^2}{f(r)} + r^2 d\Omega_2^2,
    \end{eqnarray}
where $d\Omega_2^2 = d\theta^2 + \sin^2\theta d\varphi^2$.
With this metric assumption, Equations (\ref{var eom1})--(\ref{var eom3}) read
    \begin{eqnarray}
        \label{eom00}
      r f' \!+\!  f \!+\! r^2 V \!-\!  1\!=\!0, \\
       \label{eom11}
        \left(\frac{r k'}{k}\!+\!2 r^2 \lambda  \varphi '^2 \!+\! 1\right) f
             \!+\! r^2 V \!-\! 1\!=\!0, \\
       \label{eom22}
       \left(\frac{r k'}{k}\!+\!2\right) f'
          \!+\! \left(\frac{2 r k''}{k}\!-\!\frac{r k'^2}{k^2}\!+\!\frac{2 k'}{k}\right) f
          \!+\!4 r V\!=\!0,\\
       \label{eomphi}
       \lambda\!  \left[\! \left(\!\frac{k'}{k}\!+\!\frac{4}{r}\!\right)\!f \varphi '
         \!+\!2 f\varphi ''  \!+\!f' \varphi '\!+\! U_{\varphi}\!\right]\!+\!2 f \lambda ' \varphi '\!
         +\!V_{\varphi}\!=\!0,\\
       \label{eomlambda}
       f \varphi '^2 \!-\! U\!=\!0,
    \end{eqnarray}
where the primes denote the derivatives with respect to the coordinate $r$.
Equations~\eqref{eom00} and \eqref{eomlambda} give, respectively, the solutions of the scalar potentials $V(\varphi(r))$ and $U(\varphi(r))$ as functions of $r$ after the metric function $f(r)$ and the scalar filed $\varphi(r)$ are known:
    \begin{eqnarray}
       V&=&\frac{1-f-r f'}{r^2},  \label{Veq00}\\
       U&=&f \varphi '^2.  \label{Ueqlambda}
    \end{eqnarray}

Substituting the above solution \eqref{Veq00} into Equations~\eqref{eom11} and \eqref{eom22}, we get the solution of the Lagrange multiplier
    \begin{eqnarray}
       \lambda&=&\frac{1}{2 r \varphi '^2} \left( \frac{f'}{f}-\frac{k'}{k} \right)
    \end{eqnarray}
and the relationship between $k(r)$ and $f(r)$:
    \begin{eqnarray}
       \left(\frac{k'}{k} \!-\! \frac{2}{r}\right) f'
        \!+\! \left(\frac{2 k''}{k} \!-\! \frac{k'^2}{k^2}
        \!+\!\frac{2 k'}{r k} \!-\! \frac{4}{r^2}\right) f \!+\!  \frac{4}{r^2}=0. \label{relationship_f_k}
    \end{eqnarray}

Then, considering $V_{\varphi} = V'/\varphi'$, $U_{\varphi} = U'/\varphi'$ and substituting Equations~\eqref{Veq00}--\eqref{relationship_f_k} into Equation~\eqref{eomlambda}, one can easily show that the equation of motion of the scalar field $\varphi(r)$~\eqref{eomlambda} is satisfied automatically. Therefore, there are only four independent field equations, e.g., Equations~\eqref{Veq00}--\eqref{relationship_f_k}. Usually, by giving the expressions of the scalar potential $U(\varphi)$ and $V(\varphi)$, we can solve all of the field equations. However, it is very hard to obtain an analytic solution via this method. Note that the field equation for the function $k(r)$ is of {the second order}, once $k(r)$ and $\varphi(r)$ are given, we could get the analytic solution for $f(r)$, $\lambda(r)$, $U(\varphi)$ and $V(\varphi)$.

\section{Solutions}\label{solution}

In this section, we look for an analytical solution with asymptotic flat spacetime for the case of a constant Lagrange multiplier.
Our solution is given by
\begin{eqnarray}
         k(r) \!\!&=&\!\!  e^{-{s}/{r^3}} f(r)
                    , \label{k(r)} \\
         f(r) \!\!&=&\!\! \frac{4 r^2}{9 s}
                \Big[
                     \mathcal{F}_1(r) \!-\!\mathcal{F}_2(r)
                \Big]    , \label{f(r)} \\
         \varphi(r) \!\!&=&\!\!  \sqrt{-\frac{2 s}{3 \lambda  r^3}},\\
        U(\varphi) \!\!&=&\!\! \frac{2(\mathcal{F}_2(\Psi) \!-\! \mathcal{F}_1(\Psi))}
                                 {3 \lambda  \Psi^3}
               , \\
        V(\varphi) \!\!&=&\!\!  \frac{3 s \Psi
                         \!+\! 4  \left(s \!-\! \Psi^3\right)
                          \mathcal{F}_1(\Psi)
                         \!-\! 2\left(s \!-\! 2\Psi^3\right)
                          \mathcal{F}_2(\Psi)}{3 s \Psi^3}
                  ,
\end{eqnarray}
where $\lambda<0$, $s$ is a positive scalar parameter, and
\begin{eqnarray}
   \!\!\!\mathcal{F}_n(x) \!\!&=&\!\! e^{\frac{s}{nx^3}} \!\!
                       \left[ \!\sqrt[3]{\frac{s}{n}} \!\left(
                             \!\Gamma \Big(\!-\!\frac{1}{3},\frac{s}{nx^3}\Big)\!
                             -\! \Gamma \!\Big(\!-\frac{1}{3}\Big)\!
                             \right)\!
                             -\! 9 M
                       \right]\!,~~~~\\
   \Psi \!\!&=&\!\! \sqrt[3]{-\frac{2 s}{3 \lambda  \varphi ^2}}.
\end{eqnarray}

\begin{figure}
\includegraphics[width=6cm,height=4cm]{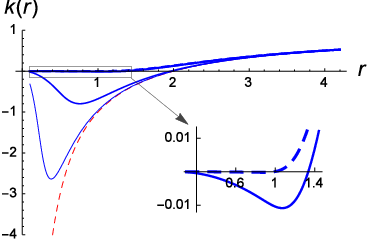}
\includegraphics[width=6cm,height=4cm]{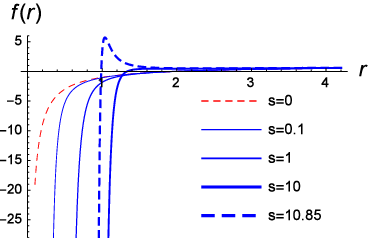}
  \caption{The shapes of the metric functions $k(r)$ and $f(r)$ in (\ref{k(r)}) and (\ref{f(r)}) compared with the solution of the Schwarzschild solution ($s=0$).
  The parameters are set to $M=1$ and $s=0$ (dashed red lines), 0.1 (the most thin line), 1, 10, and 10.85 (the most thick line).}\label{Figfrkr}
\end{figure}

Here, $\Gamma(r)$ is the Euler gamma function and $\Gamma(a,r)$ the incomplete gamma function.
Note that the Lagrange multiplier $\lambda$ does not affect the metric functions directly in the above solution.
The shapes of the metric functions $k(r)$ and $f(r)$ are shown in Figure \ref{Figfrkr}. Other static spherically symmetric black hole or  wormhole solutions with $U(\varphi)=-1$ can be found in~\cite{Myrzakulov2015}.

For {a small} scalar reduced parameter $\hat{s}\equiv s/M^3 \ll 1$,  we have
\begin{eqnarray}
       k(r) &=& 1-\frac{2 M}{r}-\left(\frac{1}{10 r^3}-\frac{M}{2 r^4}\right) {s}\nonumber \\
                    &-&\left(\frac{1}{160 r^6}+\frac{M}{12 r^7}\right)s^2
                    +\mathcal{O}\big( \hat{s} ^3\big),\\
       f(r) &=& 1-\frac{2 M}{r}+\left(\frac{9}{10 r^3}-\frac{3M}{2 r^4}\right) {s}\nonumber \\
                    &+& \left(\frac{63}{160 r^6}-\frac{7 M}{12 r^7}\right) s^2
                    +\mathcal{O}\big( \hat{s} ^3\big),
\end{eqnarray}
{which show that} the above black hole solution will become the Schwarzschild one when the scalar parameter vanishes.
The relation of the parameter $M$ (the mass of the black hole, see the discussion later) and the horizon radius of the black hole is given by
\begin{eqnarray}
M&=&\frac{\sqrt[3]{s}}{18 \left(e^{\frac{s}{2 r_{\text{h}}^3}}-1\right)}
 \left[ \Gamma \Big(-\frac{1}{3}\Big) \left(2^{2/3}-2 e^{\frac{s}{2 r_{\text{h}}^3}}\right)
 \right. \nonumber \\
 &+&  \left. 2 e^{\frac{s}{2 r_{\text{h}}^3}} \Gamma \Big(-\frac{1}{3},\frac{s}{r_{\text{h}}^3}\Big)
            -  2^{2/3} \Gamma \Big(-\frac{1}{3},\frac{s}{2 r_{\text{h}}^3}\Big)
            \right],
  \label{M_rh_exact}
\end{eqnarray}
which can be approximated as
\begin{eqnarray}
M=  \frac{r_{\text{h}}}{2}  \left(1+\frac{3 s}{20 r_{\text{h}}^3}-\frac{s^2}{96 r_{\text{h}}^6} + \mathcal{O}\big( (s/r^3_{\text{h}}) ^3\big)\right)
  \label{M_rh}
\end{eqnarray}
for the small $s/r^3_{\text{h}}$. Note that, since $M=r_{\text{h}}/2$ in the lowest order of $s/r^3_{\text{h}}$, we can replace $\mathcal{O}\big( (s/r^3_{\text{h}}) ^3\big)$ in the above expression with $\mathcal{O}\big( \hat{s}^3\big)$.
Hereinafter, we omit the term $\mathcal{O}\big( \hat{s} ^3\big)$.
From the above expression (\ref{M_rh}), we can get the approximate solution of the horizon radius
\begin{eqnarray}
r_{\text{h}} 
   =2 M \left(1-\frac{3 s}{160 M^3}-\frac{83 s^2}{153600 M^6}\right).
   \label{rh_M}
\end{eqnarray}

For simplicity, we define a new dimensionless parameter
\begin{eqnarray}
\bar{s}=\frac{3 \hat{s}}{160}=\frac{3 s}{160M^3} \label{bars}
\end{eqnarray}
and rewrite the horizon radius as
\begin{eqnarray}
r_{\text{h}} =2 M \left( 1 -\bar{s} -\frac{83}{54}\bar{s}^2\right). \label{HorizonRadius}
\end{eqnarray}

\begin{figure}
\includegraphics[width=6cm,height=4cm]{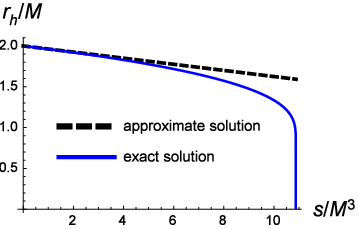}
\includegraphics[width=6cm,height=4cm]{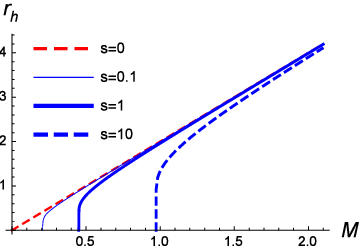}
  \caption{The relations between the horizon radius $r_{\text{h}}$ and the scalar parameter $s$ and the mass $M$.
  \textbf{Left}: $r_{\text{h}}\sim s$. The black dashed line and blue line denote the approximate solution (\ref{HorizonRadius}) (for small $\hat{s}$) and precise numerical solution, respectively.  \textbf{Right}: $r_{\text{h}}\sim M$. The red dashed line corresponds to the case of the Schwarzschild black hole case.}\label{rhMSrelation}
\end{figure}

The relation between the horizon radius $r_{\text{h}}$ and the scalar parameter $s$ is shown in Figure \ref{rhMSrelation}, from which it can be seen that the expression (\ref{HorizonRadius}) is accurate for $\hat{s} < 4$. Especially, the horizon radius will {decrease slowly first and then rapidly with the increasing of the scalar parameter.}
Our result shows that {the $\lambda\varphi$ fluid} will decrease the horizon radius of the black hole. This could be understood as the {attractive effect} of {the $\lambda\varphi$ fluid}, which is opposite to the repulsive effect of dark energy driving the expansion of the universe \cite{MaChenSun:2018NPB}.

The asymptotic behaviors at spatial infinity $r \rightarrow \infty$ and at origin $r \rightarrow 0$ are, respectively,
\begin{eqnarray}
   k(r\rightarrow\infty) &=& 1-\frac{2 M}{r}
                            -\frac{s}{10 r^3}
                            +\mathcal{O}\Big(({M}/{r})^4\Big), \label{asymptoticInfinity_k}\\
   f(r\rightarrow\infty) &=& 1-\frac{2 M}{r}
                            +\frac{9s}{10 r^3}
                            +\mathcal{O}\Big(({M}/{r})^4\Big), \label{asymptoticInfinity_f}
\end{eqnarray}
and
\begin{eqnarray}
   k(r\rightarrow 0) &\rightarrow& -\frac{4r^2}{9s}
                        \left(9 M+\sqrt[3]{s} \,\Gamma \left(-{1}/{3}\right)\right)
                        ,
                        \label{asymptotic_r_0_k}\\
   f(r\rightarrow 0) &\rightarrow& -\frac{4 r^2}{9 s}
                         \left(9 M+\sqrt[3]{s}\,\Gamma \left(-{1}/{3}\right)\right) e^{\frac{s}{r^3}}
                         .
                        \label{asymptotic_r_0_f}
\end{eqnarray}

From the last two expressions, one can see that $k(r\rightarrow 0)\rightarrow 0$ and $f(r\rightarrow0)\rightarrow -\infty$ for $9 M+\sqrt[3]{s}\,\Gamma \left(-{1}/{3}\right)>0$, {which is confirmed} from Figure \ref{Figfrkr}. Therefore, for a given mass $M$, the condition for having an event horizon is
\begin{eqnarray}
   0 \leq s < s_{\text{max}}\equiv \left(\frac{9M}{-\Gamma \left(-{1}/{3}\right)}\right)^3
       \thickapprox10.8741 M^3. \label{Condition_S}
\end{eqnarray}

The left figure in Figure~\ref{rhMSrelation} shows that the horizon radius approaches its maximum $r_{\text{h}}=2M$ and minimum $r_{\text{h}}=0$ when the scalar parameter approaches its minimum $s=0$ and its maximum $s=s_{\text{max}}$, respectively. On the other hand, for a fixed scalar parameter $s$, the condition (\ref{Condition_S}) becomes
\begin{eqnarray}
   M  \geq  M_{\text{min}}(s)\equiv -\frac{1}{9} \Gamma \big(-{1}/{3}\big) s^{1/3}
     =0.451373 ~s^{1/3}, \label{Condition_M}
\end{eqnarray}
which is different from the case of the Schwarzschild solution for nonvanishing $s$. The right figure in Figure~\ref{rhMSrelation} shows that the horizon radius decreases linearly with the decrease of the mass $M$ for large $M$, i.e., $r_{\text{h}}\simeq 2M$. However, when the mass approaches to its minimum given in~(\ref{Condition_M}), the horizon radius drops quickly to zero.

The invariant of the combinations of the Riemann curvature, the Kretschman scalar, for the small $\hat{s} \ll 1$ is
\begin{eqnarray}
R_{\mu\nu\lambda\rho}R^{\mu\nu\lambda\rho}=\frac{48 M^2}{r^6}+\frac{24  (2 M-r)M s}{r^9}
           +\mathcal{O}\big( \hat{s} ^2\big),
\end{eqnarray}
where the first term on the right-hand side is the result of the Schwarzschild solution.
However, note that the singularity behavior of the above invariant at $r \rightarrow 0$
\begin{eqnarray}
R_{\mu\nu\lambda\rho}R^{\mu\nu\lambda\rho}
  \rightarrow \frac{16 }{3} \left(9 M+\sqrt[3]{s} \, \Gamma  \left(-{1}/{3}\right)\right)^2
   \frac{e^{{2 s}/{r^3}}}{r^6} \label{Rabcd2rto0}
\end{eqnarray}
is very different from that of the Schwarzschild black hole
\begin{eqnarray}
R_{\mu\nu\lambda\rho}R^{\mu\nu\lambda\rho}
  \rightarrow \frac{48 M^2}{r^6}. \label{Rabcd2}
\end{eqnarray}

\begin{figure}
\includegraphics[width=6cm,height=4cm]{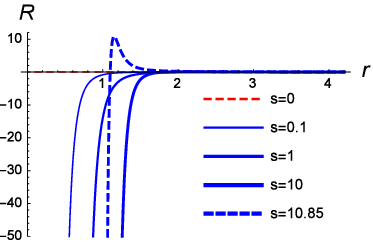}
\includegraphics[width=6cm,height=4cm]{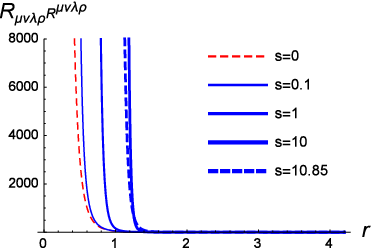}
  \caption{The shapes of the scalar curvature $R$ and the Kretschman scalar $R_{\mu\nu\lambda\rho}R^{\mu\nu\lambda\rho}$ compared with that of the Schwarzschild solution.   The mass parameter is set to $M=1$. }\label{FigRr}
\end{figure}

It is clear that there is another singularity factor $e^{{2 s}/{r^3}}$ in (\ref{Rabcd2rto0}) besides ${1}/{r^6}$.
The shapes of the scalar curvature $R$ and $R_{\mu\nu\lambda\rho}R^{\mu\nu\lambda\rho}$ are shown in Figure \ref{FigRr}.

Next, we analyze the spacetime structure of the black hole solution. For convenience, we let
\begin{eqnarray}
         k(r) = \psi(r) f(r),
\end{eqnarray}
where
\begin{eqnarray}
         \psi(r) =  e^{-{s}/{r^3}}.
\end{eqnarray}

In order to analyze the causal structure of the black hole, we calculate the metric in Kruskal--Szekeres coordinates. First, we define the tortoise coordinate $r^{*}$:
\begin{eqnarray}
         d r^{*}= \frac{dr} {\varphi(r)}, \label{tortoiseCoordinate}
\end{eqnarray}
where $\varphi(r)=\sqrt{\psi(r)} f(r)$. In order to integrate the above equation, we expand the function $\varphi(r)$ around the horizon to {the first order}:
\begin{eqnarray}
       \varphi(r) =  \varphi'(r_{\text{h}})(r-r_{\text{h}}).
        \label{PhirAroundrh}
\end{eqnarray}

Here, we have used the fact $\varphi(r_{\text{h}})=0$. Note that the parameter $\varphi'(r_{\text{h}})$ is a function of $M$ and $s$. Therefore, when $r \rightarrow r_{\text{h}}$, the relation between $r^{*}$ and $r$ has the following form:
\begin{eqnarray}
          r^{*} = \frac{\ln\big(r/r_{\text{h}} -1\big)}{\varphi'(r_{\text{h}})} \rightarrow -\infty,
           \label{r*Aroundrh}
\end{eqnarray}
which shows that the surface at the horizon has been pushed to infinity in the tortoise coordinate.
For the case of $s=0$, we have ${\varphi'(r_{\text{h}})} = 1/r_{\text{h}} =1/2M$. When $r \rightarrow +\infty$, the relation becomes
\begin{eqnarray}
          r^{*} = r+ 2M\,{\ln\big(r/2M -1\big)} \rightarrow +\infty.
\end{eqnarray}

With the coordinate translation (\ref{tortoiseCoordinate}), the metric (\ref{metric1}) is given by
    \begin{eqnarray}
        \label{metric2}
       ds^2=\psi f \big( -dt^2+ dr^{*2}\big) + r^2 d\Omega^2.
    \end{eqnarray}

It is clear that the causal structure outside the horizon, which is described by $dt=\pm \frac{dr} {\varphi(r)} =\pm dr^{*}$, is almost the same as the Schwarzshild black hole for the case of $\hat{s}\ll 1$. The tortoise coordinate is only sensibly related to the coordinate $r$ when $r\ge r_{\text{h}}$. Therefore, we introduce coordinates $u$ and $v$:
    \begin{eqnarray}
       u &=& t-r^{*},  \label{EFcoordinate_u}\\
       v &=& t+r^{*}.\label{EFcoordinate_v}
    \end{eqnarray}

It can be seen that $u=$constant and $v$=constant denote outgoing and ingoing radial null geodesics, respectively. In terms of Eddington--Finkelstein coordinates (\ref{EFcoordinate_u}) and (\ref{EFcoordinate_v}), the metric (\ref{metric2}) is further rewritten as
    \begin{eqnarray}
        \label{metricEFcoordinate1}
       ds^2=-\psi f du^2 - \sqrt{\psi} \big( dudr+ drdu\big) + r^2 d\Omega^2,
    \end{eqnarray}
or
    \begin{eqnarray}
        \label{metricEFcoordinate2}
       ds^2=-\psi f dv^2 + \sqrt{\psi} \big( dvdr+ drdv\big) + r^2 d\Omega^2.
    \end{eqnarray}

The radial null curves are given by
    \begin{eqnarray}
        \frac{du}{dr}=\left\{\begin{array}{c}
                        0 \\
                        -\frac{2}{\varphi(r)}
                      \end{array} \right.
        ,
    \end{eqnarray}
or
    \begin{eqnarray}
        \frac{dv}{dr}=\left\{\begin{array}{c}
                        0 \\
                        \frac{2}{\varphi(r)}
                      \end{array} \right.
        .
    \end{eqnarray}

In the ($v$, $r$) coordinates, one can cross the event horizon on future-directed curves, while in the ($u$, $r$) coordinates, one can pass though the event horizon along past-directed ones. In fact, spacetime has been extended in two different directions, the future and the~past.

In the ($u$, $v$) coordinates, the metric reads
    \begin{eqnarray}
       ds^2=-\psi f dudv  + r^2 d\Omega^2,
    \end{eqnarray}
and the horizon $r=r_{\text{h}}$ is infinitely far away and it is at either $u=+\infty$ or $v=-\infty$.
We can pull them into finite positions by the following transformation:
    \begin{eqnarray}
        \bar{u} &=& -e^{-\varphi'(r_{\text{h}}){u}/{2} } = -e^{\varphi'(r_{\text{h}})(r^* -t )/{2} } ,\\
        \bar{v} &=& ~~e^{\varphi'(r_{\text{h}}){v}/{2}} ~~ =  \,~~e^{\varphi'(r_{\text{h}})(r^* +t)/{2} } .
    \end{eqnarray}
Now the metric becomes
    \begin{eqnarray}
       ds^2=-\mathcal{F}(r) d\bar{u}d\bar{v}   + r^2 d\Omega^2,
    \end{eqnarray}
where
    \begin{eqnarray}
       \mathcal{F}(r)=\frac{4 k(r) }{\big(\varphi'(r_{\text{h}})\big)^2}  e^{-\varphi'(r_{\text{h}})r^*}.
    \end{eqnarray}

Next, we transform the null coordinates $\bar{u}$ and $\bar{v}$ to one timelike coordinate $\mathcal{T}$ and one spacelike coordinate $\mathcal{R}$:
    \begin{eqnarray}
       \mathcal{T}&=&\frac{1}{2} \big( \bar{v} + \bar{u} \big)
                   = e^{\varphi'(r_{\text{h}}) r^*/2}  \sinh \big( \varphi'(r_{\text{h}}) t/2 \big),\\
       \mathcal{R}&=&\frac{1}{2} \big( \bar{v} - \bar{u} \big)
                   =e^{\varphi'(r_{\text{h}})r^*/2}  \cosh \big( \varphi'(r_{\text{h}}) t/2 \big).
    \end{eqnarray}

Then, the metric in Kruskal--Szekeres coordinates becomes
    \begin{eqnarray}
       ds^2=\mathcal{F}(r) \left( -d\mathcal{T}^2 + d\mathcal{R}^2 \right)   + r^2 d\Omega^2,
    \end{eqnarray}
where $r$ is defined from
    \begin{eqnarray}
       \mathcal{T}^2 - \mathcal{R}^2 = e^{\varphi'(r_{\text{h}}) ~r^*(r)}.
    \end{eqnarray}

It is clear that the function $\mathcal{F}(r)$ is smooth for $r>0$ and $r\neq r_{\text{h}}$.
We need to know whether it is also smooth at $r=r_{\text{h}}$. To this end, we consider Equations (\ref{PhirAroundrh}) and (\ref{r*Aroundrh}) and expand $\mathcal{F}(r)$ around $r=r_{\text{h}}$:
    \begin{eqnarray}
       \mathcal{F}(r\rightarrow r_{\text{h}})=\frac{4 r_{\text{h}}  } {\varphi'(r_{\text{h}})} e^{-r_{\text{h}}\varphi'(r_{\text{h}})-s/(2r_{\text{h}}^3)} +\mathcal{O}\big( r-r_{\text{h}}\big),
    \end{eqnarray}
which shows that $\mathcal{F}(r)$ is smooth at $r=r_{\text{h}}$. Therefore, $\mathcal{F}(r)$ is a smooth function of $r$ for $r>0$. When $s\rightarrow 0$, $\varphi'(r_{\text{h}})=1/r_{\text{h}}=1/{2M}$ and hence $\mathcal{F}(r\rightarrow r_{\text{h}})=4r_{\text{h}}^2 e^{-1}$.

\begin{figure}
  \includegraphics[width=6cm,height=4.6cm]{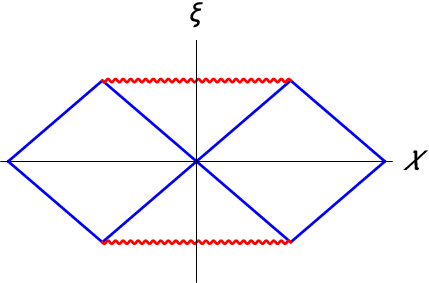}
  \caption{The Penrose--Carter diagram. }\label{PenroseDiagram}
\end{figure}

At last, with the following transformation
    \begin{eqnarray}
        \mathcal{T} +  \mathcal{R} &=& \tan (\xi + \chi),~~\\
        \mathcal{T} -  \mathcal{R} &=& \tan (\xi - \chi),
    \end{eqnarray}
we obtain the following metric
    \begin{eqnarray}
       ds^2=\mathcal{F}(r) \frac{ -d\xi^2 + d\chi^2 }{\cos^2(\xi + \chi)\cos^2(\xi - \chi)}   + r^2 d\Omega^2.
       \label{metricconformal}
    \end{eqnarray}
Note that the range of the coordinates $\xi$  and $\chi$ is $-\pi/2<\xi + \chi<\pi/2$, $-\pi/2 < \xi - \chi<\pi/2$, and $-\pi/4<\xi<\pi/4$. From the metric (\ref{metricconformal}), we draw the Penrose--Carter diagram (also called conformal diagram) in the $(\xi,\chi)$ coordinates for the black hole solution in Figure \ref{PenroseDiagram}, which is similar to the Schwarzschild case.

\section{Black Hole Mass, Temperature and Entropy}\label{TemperatureEntropy}

{In this section, we calculate the black hole mass, temperature, and entropy.}
With the above asymptotic behaviors (\ref{asymptoticInfinity_k}) and (\ref{asymptoticInfinity_f}) of the metric functions $k$ and $f$, we can calculate the Komar integral associated with the timelike Killing vector $K^{\mu }=(1,0,0,0)$, i.e., the total energy of the corresponding static spacetime,
\begin{eqnarray}
E_R &=& \frac{1}{4 \pi}\int _{\partial \Sigma }\sqrt{\gamma ^{(2)}}d^2x ~
            n_{\mu }\sigma _{\nu }\nabla ^{\mu }K^{\nu }  \nonumber \\
    &=& \left. \frac{r^2 \sqrt{f(r)} k'(r)}{2 \sqrt{k(r)}} \right|_{r\to \infty } \nonumber  \\
    &=& M.
\end{eqnarray}
Therefore, the parameter $M$ is the mass of the black hole. The relation of the black hole mass and the horizon radius is given in Equations~(\ref{M_rh}) and (\ref{rh_M}).

Next, we calculate the Hawking temperature of the black hole. From the formula
\begin{eqnarray}
T_{\text{H}} &=&\left.  \frac{1}{2\pi} \sqrt{-\frac{1}{2} (\nabla_{\mu} \xi_{\nu} ) (\nabla^{\mu} \xi^{\nu} )  }
        \right|_{\Sigma} \nonumber \\
     &=&    \left.  \frac{1}{4\pi}\sqrt{-g^{tt} g^{rr} (\partial_r g_{tt})^2}
        \right|_{\Sigma} \nonumber \\
 &=& \frac{1}{4 \pi } \left. {\sqrt{\psi(r_{\text{h}})} \; \partial_r f} \right|_{\Sigma}\,\,\,\,\, ,
\end{eqnarray}
where $\xi_{\mu}$ is a Killing vector on the Killing horizon $\Sigma$,
we can calculate the Hawking temperature with the existence of {the $\lambda\varphi$ fluid}
\begin{eqnarray}
T_{\text{H}} =
     \frac{\sqrt[3]{s} ~{Y}}{12 \pi  r_{\text{h}}^2 X} \,\,  ,
\label{HawkingTemperature0}
\end{eqnarray}
where
\begin{eqnarray}
X &=&  1-e^{-\frac{s}{2 r_{\text{h}}^3}}, \label{X} \\
Y &=&  \left(2^{2/3}-2\right) \Gamma \left(-\frac{1}{3}\right)
          \nonumber \\
     &-&  2^{2/3} \Gamma \left(-\frac{1}{3},\frac{s}{2 r_{\text{h}}^3}\right)
      +2 \Gamma \left(-\frac{1}{3},\frac{s}{r_{\text{h}}^3}\right)  .
\label{Y}
\end{eqnarray}

For small scalar parameter $s$, it becomes
\begin{eqnarray}
T_{\text{H}}&=&
  \frac{1}{4 \pi  r_{\text{h}}}\left(1-\frac{s}{20 r_{\text{h}}^3}+\frac{3 s^2}{160 r_{\text{h}}^6}\right)
  \nonumber\\
  &=&  \frac{1}{8 \pi  M} \left( 1+\frac{2}{3}\bar{s} + \frac{55}{27} \bar{s}^2  \right).
   \label{HawkingTemperature}
\end{eqnarray}

The above result shows that, compared with the Schwarzschild case $T_{\text{H}}=\frac{1}{8 \pi  M} $, {the $\lambda\varphi$ fluid} increases the Hawking temperature of the black hole.

At last, we calculate the black hole entropy. The scalar parameter $s$ should be treated as a new thermodynamic variable. Then, the black hole first law reads
\begin{eqnarray}
dM = T_{\text{H}} dS + \mathcal{A}ds,
\end{eqnarray}
where $\mathcal{A}$ is a thermodynamic quantity conjugated to $s$.
For fixed $s$, the entropy can be calculated with
\begin{eqnarray}
S &=& \int \frac{dM}{T_{\text{H}}}
  = \int \frac{1}{T_{\text{H}}}
         \left( \frac{\partial M}{\partial r_{\text{h}}}  \right)
         dr_{\text{h}} \nonumber \\
&=& \int \frac{\pi  s}{r_{\text{h}}^2 \left(e^{\frac{s}{2 r_{\text{h}}^3}}-1\right)} dr_{\text{h}}. \label{entropyS1}
\end{eqnarray}

To the second order of $\bar{s}$, the entropy is given by
\begin{eqnarray}
S = \pi  r_{\text{h}}^2
   \left(1 +\frac{s}{2 r_{\text{h}}^3}
         -\frac{s^2}{96 r_{\text{h}}^6}
   \right),  \label{entropySa}
\end{eqnarray}
or
\begin{eqnarray}
S= 4 \pi  M^2
   \left( 1 +\frac{4}{3}\bar{s}  + \frac{43}{54} \bar{s}^2
   \right). \label{entropySb}
\end{eqnarray}

It is obvious that the relation $S=A/4 =\pi  r_{\text{h}}^2$ of the Schwarzschild black hole has been modified by {the $\lambda\varphi$ fluid}.

\section{Black Hole Radiation Spectrum with {the $\lambda\varphi$ Fluid}}\label{RadiationSpectrum}

It is known that nothing can escape from the horizon of a black hole in classical general relativity. However, by considering the effect of quantum field theory, Hawking found that a positive energy particle from a pair created virtually around the black hole horizon can {escape from} the horizon through tunneling, and this process results in the famous Hawking radiation. It was shown that the black hole radiation spectrum obeys the thermal distribution \cite{Hawking1974,Hawking1975}. This will result in the black hole information paradox since the entropy will increase through the Hawking process.

In~\cite{ZhangCai2009}, the authors showed that the black hole radiation spectrum is not perfectly thermal but nonthermal if the constraint of energy conservation is introduced, and the problem of the information paradox can be  solved. In fact, there were other different schemes for this problem,
see~\cite{Aharonov1987,Wilczek1989,Bekenstein1993,Wilczek2000,Horowitz2004,Hawking2005,Lloyd2006,Braunstein2007} for examples.

In this section, we use the statistical mechanical method introduced in~\cite{MaChenSun:2018EPL,MaChenSun:2018NPB} to calculate the nonthermal black hole radiation spectrum, {which describes the statistical distribution of the radiated particles's energy.} This method is based on canonical typicality~\cite{MaChenSun:2018EPL,Tasaki1998,Goldstein2006,Popescu2006}.
We first give a brief review of this method.

The density matrix of a black hole B with mass $M$, charge $Q$ and angular momentum $J$ is given by
\begin{eqnarray}
   \rho_{\text{B}} = \sum_{i} \frac{1}{\Omega(M,Q,J)}|M,Q,J\rangle_i \langle M,Q,J| ,
\end{eqnarray}
where $|M,Q,J\rangle_i$ and ${\Omega(M,Q,J)}$ are the $i$th eigenstate and the number of microstates of the black hole, respectively. Now, we consider Hawking radiation. When particles are radiated from the black hole, the black hole system can be viewed as two parts, the radiation field R with mass $\omega$, charge $q$, and angular momentum
$j$, and the remaining black hole $\text{B}'$ with mass $M-\omega$, charge $Q-q$, and angular momentum
$J-j$. By considering the conservation of black hole ``hairs'' and tracing over all the degree of freedom of $\text{B}'$, one can obtain the density matrix of R~\cite{MaChenSun:2018EPL,MaChenSun:2018NPB}:
\begin{eqnarray}
   \rho_{\text{R}} = \sum_{\omega,q,j} p({\omega,q,j,M,Q,J})  |\omega,q,j\rangle \langle\omega,q,j| , \label{rhoR}
\end{eqnarray}
where $|\omega,q,j\rangle$ is the eigenstate of the radiation field, and the distribution probability of the radiation is turned out to be ~\cite{MaChenSun:2018EPL,MaChenSun:2018NPB}
\begin{eqnarray}
   p({\omega,q,j,M,Q,J}) = \exp \left(-\delta S_{\text{BB}'}({\omega,q,j,M,Q,J}) \right), \label{probability}
\end{eqnarray}
with the entropy difference between B and $\text{B}'$  given by
\begin{eqnarray}
   \delta S_{\text{BB}'} = S_{\text{B}}(M,Q,J) - S_{\text{B}'} (M-\omega,Q-q,J-j). \label{deltaS}
\end{eqnarray}

We can calculate the radiation spectrum of a black hole with the above expressions (\ref{rhoR})--(\ref{deltaS}).

The corrected radiation spectrum of a Schwarzschild black hole is a function of the black mass or the horizon
radius:
\begin{eqnarray}
   p(\omega,M) &=& e^{-\pi \left[  r_{\text{h}}^2(M) -r_{\text{h}}^2(M-\omega) \right]} \nonumber \\
       &=& e^{-8\pi M\omega+ 4\pi\omega^2 }, \label{radiationSpectrumSchwarzschild}
\end{eqnarray}
which is in accord with the result derived through the quantum tunneling method~\cite{Wilczek2000}.
The corrected term $4\pi\omega^2$ comes from the higher order of the energy $\omega$.

For the black hole with {the $\lambda\varphi$ fluid} considered in this paper,
the radiation spectrum can be calculated with
\begin{eqnarray}
   p(\omega,M) = e^{-\delta S(\omega,M)},
     \label{radiationSpectrum}
\end{eqnarray}
where $\delta S(\omega,M) =S(M)-S(M-\omega)$. Considering the expressions of the entropy (\ref{entropyS1}), temperature (\ref{HawkingTemperature0}), and mass (\ref{M_rh_exact}) and keeping to the second order of $\omega$, we have
\begin{eqnarray}
   p(\omega,M) = e^{- \beta_{\text{H}} \omega  + \chi \omega ^2  },
     \label{radiationSpectrum}
\end{eqnarray}
where
\begin{eqnarray}
   \beta_{\text{H}} &=& \frac{\partial S}{\partial M} =\frac{1}{T_{\text{H}}}
      = \frac{12 \pi  r_{\text{h}}^2 X}
             {\sqrt[3]{s} ~{Y}}  ,  \label{betaH}\\
   \chi &=& \frac{1}{2} \frac{\partial^2 S}{\partial M^2} = \frac{1}{2} \frac{\partial \beta_{\text{H}}}{\partial M}
     = \frac{1}{2} \frac{\partial_{r_{\text{h}}} (1/T_{\text{H}})}{\partial_{r_{\text{h}}} M} \nonumber \\
    &=& \frac{432 \pi  r_{\text{h}}^6 X^4}
             {s^2 Y^3}
       +\frac{144 \pi  r_{\text{h}}^5 e^{\frac{s}{2r_{\text{h}}^3}} X^3}
             {s^{5/3} Y^2 }
       -\frac{108 \pi  r_{\text{h}}^2 X^2}
             {s^{2/3} Y^2}. ~~~~  \label{chi}
\end{eqnarray}

To {the second order} of $\bar{s}$,
\begin{eqnarray}
   \beta_{\text{H}} &=&  8 \pi  M   \left(1 -\frac{2}{3}\bar{s} -\frac{43}{27} \bar{s}^2 \right),  \label{betaH2}\\
   \chi &=& 4\pi   \left(1+\frac{4}{3}\bar{s}  +\frac{215}{27} \bar{s}^2 \right). \label{chi2}
\end{eqnarray}

\begin{figure}
  \includegraphics[width=6cm]{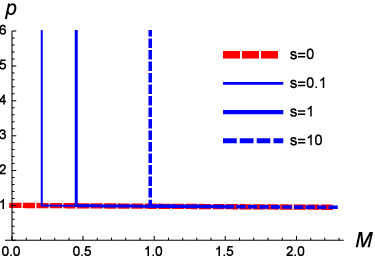}
  \caption{The radiation spectrum (\ref{radiationSpectrum})--(\ref{chi}) as a function of the black hole mass $M$ for different values of the scalar parameter $s$. The parameters are set to $s=0,~0.1,~1,~10$ and $\omega=10^{-3}$.  } \label{Figprhs}
\end{figure}

The radiation spectrum as a function of the black hole $M$ is plotted in Figure~\ref{Figprhs} for different values of the scalar parameter $s$ and $\omega=10^{-3}$. The black hole mass will decrease with the Hawking radiation. It can be seen that the Hawking radiation will be accelerated rapidly at its late stage when the black hole mass approaches to its minimum $M_{\text{min}}(s)$, which is given by (\ref{Condition_M}) and takes the values of $0.21,~0.45,~0.97$ for $s=0.1,~1,~10$, respectively.

Next, we investigate the evaporation process of the black hole {with the $\lambda\varphi$ fluid} based on the radiation spectrum. We mainly calculate the lifetime of the black hole and discuss the effect of the scalar parameter on it.
The result (\ref{HawkingTemperature}) shows that {the $\lambda\varphi$ fluid} increases the Hawking temperature of the black hole, and hence makes the Hawking radiation hotter. The result is opposite to the dark energy case, which is coincident with the fact that the horizon radius of the black hole is reduced by {the $\lambda\varphi$ fluid}, but enlarged by the dark energy. For the case of black hole with {the $\lambda\varphi$ fluid} (dark energy), decreasing (enlarging) of the black hole horizon $r_{\text{h}}$ will lead to enlarging (decreasing) of the black hole surface gravity $g_{\text{h}}=M/r_{\text{h}}^2$, and hence the Hawking temperature $T_{\text{H}}=g_{\text{h}}/2\pi$ will become hotter (lower).

In~\cite{Sendouda2006}, Sendouda investigated the Hawking radiation of five-dimensional small primordial black holes in the Randall--Sundrum braneworld. It was found that the Hawking temperature of a black hole will be reduced by the large extra dimension and the spectra of emitted particles via Hawking radiation are drastically changed. In~\cite{Dai2008}, Dai considered the nonrotating black hole in braneworld model and showed that, for the brane with nonzero tension, the horizon radius of the black hole increases with brane tension and hence the brane tension lowers the Hawking temperature of the black hole and the average energy of the emitted particles. These effects are also opposite to our case with {the $\lambda\varphi$ fluid}.

\begin{figure}
  \includegraphics[width=6cm]{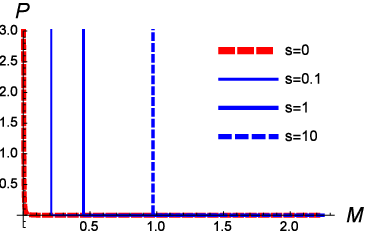}
  \caption{The radiation power  $P$ as a function of the black hole mass $M$ for different values of the scalar parameter $s$. The parameters are set to $s=0,~0.1,~1,~10$.  } \label{FigPowerMs}
\end{figure}

With the expression (\ref{HawkingTemperature}) of the Hawking temperature, we can calculate the lifetime of the black hole. By following the Stefan--Boltzmann power law, we can write the radiation power of the black hole as a function of the Hawking temperature
\begin{eqnarray}
   P = \sigma A_{\text{h}} T_{\text{H}}^4 = \frac{s^{4/3} Y^4}{15\times 12^4 \, \pi  r_{\text{h}}^6 X^4}, \label{RadiationPower}
\end{eqnarray}
where $\sigma=\pi^2/60$ is the Stefan constant, $A_{\text{h}}=4\pi r_{\text{h}}^2$ is the area of the horizon, and $X$ and $Y$ are given by Equations (\ref{X}) and (\ref{Y}), respectively. The radiation power (\ref{RadiationPower}) as a function of the black hole
mass $M$ for different values of the scalar parameter $s$ is plotted in Figure~\ref{FigPowerMs}, which shows that the radiation power will be increased rapidly at the late stage of the Hawking radiation. This is in accordance with the result given in Figure~\ref{Figprhs}.
By considering the energy conservation law for the black hole, we have
\begin{eqnarray}
   \frac{d M}{dt} + P =0,
\end{eqnarray}
which can be written as
\begin{eqnarray}
   -\frac{d M}{dt}
   = \frac{ \left( 1 -\bar{s} -\frac{83}{54}\bar{s}^2\right)^2
              \left( 1+\frac{2}{3}\bar{s} + \frac{55}{27} \bar{s}^2  \right)^4}
       {15\times 2^{10} \pi M^2},
\end{eqnarray}
where we have used the results of (\ref{HorizonRadius}) and (\ref{HawkingTemperature}).
Thus, the lifetime of the black hole evaporating its mass from $M$ to $M_{\text{min}}(s)$ can be calculated by integrating the following equation:
\begin{eqnarray}
   t= -\!\!\int_{M}^{M_{\text{min}}(s)}  \!\!\!{15\times 2^{10} \pi M^2}
      \left( 1 - \frac{2}{3}\bar{s} -\frac{80}{27} \bar{s}^2 \right) {d M}.
\end{eqnarray}

Note that the upper limit is $M_{\text{min}}(s)$ rather than 0, since after {the black hole} evaporates its mass from $M$ to $M_{\text{min}}(s)$, it is not a black hole anymore.
The result is
\begin{eqnarray}
   t &=& 5120 \pi  \left(M^3-M_{\min }^3(s)\right) - 64 \pi  s \ln \left(\frac{M^3}{M^3_{\min }(s)}\right)
     \nonumber \\
   &&+ \frac{16}{3} \pi  s^2 \left(\frac{1}{M^3}-\frac{1}{M_{\min }^3(s)}\right).
\end{eqnarray}

Consider the expressions (\ref{Condition_M}) and (\ref{bars}), we have
\begin{eqnarray}
   t = 5120 \pi M^3 \left(1 - \varepsilon \bar{s}+ \frac{80}{27}\bar{s}^2\right),
\end{eqnarray}
where
\begin{eqnarray}
\varepsilon &=& \frac{2}{3} \ln \left(\frac{-2187}{160 \bar{s} \Gamma^3 \left(-\frac{1}{3}\right)}\right)
         -\frac{160 \Gamma^3 \left(-\frac{1}{3}\right)}{2187}
      -\frac{81}{2 \Gamma^3 \left(-\frac{1}{3}\right)}
    \nonumber \\
&=& 4.45 -\frac{2}{3} \ln (\bar{s}).
\end{eqnarray}

Note that the corrected term $\varepsilon \bar{s}$ is positive. Therefore, {the $\lambda\varphi$ fluid} with small scalar parameter ($\bar{s} \ll 1$) will speed up the Hawking radiation process, and hence will reduce the lifetime of the black hole.

\section{Dark Information Reduced by {the $\lambda\varphi$ Fluid}}
\label{DarkInfor}

It has been proved that the nonthermal radiation is the origin of the information correlation between the emissions radiated out from the black hole's horizon \cite{ZhangCai2009}. Different from the Hawking radiation, this information correlation can not be measured locally. Such information stored in correlation is called dark information, which was proposed to resolve the problem of the black hole information paradox \cite{ZhangCai2009,MaChenSun:2018EPL,Dong2014}, since the total information of the black hole system is conserved by considering the dark information caused by the noncanonical statistic behavior of the Hawking radiation.

Let us consider two radiated particles $a$ and $b$ that escape from the black hole horizon. Their energy distributions are not independent of each other because of the nonthermal radiation spectrum. Therefore, there are correlation between the two particles. Now, if we use two detectors to detect the two particles separately, then the correlation is hidden and cannot be probed locally. This correlation information can be detected only through the coincidence measurement of the two detectors. Therefore, the information is dark.

In this section, we study the effect of {the $\lambda\varphi$ fluid} on this correlation.

The correlation between the two nonindependent events $a$ and $b$ can be described with the mutual information \cite{Cover2012}
\begin{eqnarray}
   I(a,b)= \sum_{a,b}  p_{a,b} \ln \left( \frac{p_{a,b}}{p_a p_b} \right). \label{Iab}
\end{eqnarray}

Here, $p_a$($p_b$) is the probability for the event $a$($b$), and $p_{a,b}$ the joint probability of $a$ and $b$. For two independent events, one has $p_{a,b} = p_a p_b$ and so $I (a, b) = 0$, i.e., the mutual information vanishes. If the radiation spectrum of a black hole is perfectly thermal, just as the one found by Hawking, then the mutual information among radiations vanishes and so there is no dark information.

Now, we consider two events of the radiation process of particles $a$ and $b$ with energy $\omega_a$ and $\omega_b$, respectively. According to (\ref{radiationSpectrum}), the probability for each particle and the joint probability are given by
\begin{eqnarray}
    p_a &=&  e^{-\beta_H \omega_a+\chi \omega_a^2}, \label{pa}\\
    p_b &=&  e^{-\beta_H \omega_b+\chi \omega_b^2}, \label{pb}\\
    p_{a,b} &=& 
            e^{-\beta_H (\omega_a+\omega_b)+\chi (\omega_a+\omega_b)^2},\label{pab}
\end{eqnarray}
where $\beta_{\text{H}}$ and $\chi$ are given by Equations~(\ref{betaH}) and (\ref{chi}), respectively.
Substituting the above expressions (\ref{pa})--(\ref{pab})  into (\ref{Iab}), we have
\begin{eqnarray}
   I(a,b)= 2\chi \sum_{a,b} p_{a,b} \omega_a \omega_b. \label{Iab2}
\end{eqnarray}

By considering the relation $p_{a,b} = p(\omega_a,M) p(\omega_b,M-\omega_a)$ and doing the replacement $M-\omega_a\rightarrow M'$, we can rewrite Equation (\ref{Iab2}) as
\begin{eqnarray}
   I(a,b) &=& 2\chi \left(\sum_{\omega_a=0}^{M} p(\omega_{a},M) \omega_a \right)
                 \left(\sum_{\omega_b=0}^{M'} p(\omega_b, M') \omega_b \right)\nonumber\\
          &=& 2\chi E_a E_b,     \label{Iab3}
\end{eqnarray}
where $E_a$ and $E_b$ are, respectively, the internal energy of particles $a$ and $b$:
\begin{eqnarray}
   E_a &=& \langle \omega_a \rangle= \sum_{\omega_a=0}^{M} p(\omega_{a},M) \omega_a, \\
   E_b &=& \langle \omega_b \rangle= \sum_{\omega_b=0}^{M'} p(\omega_b, M') \omega_b.
\end{eqnarray}

For the radiation process of the black hole with {the $\lambda\varphi$ fluid}, the explicit form of the dark information is given by
\begin{eqnarray}
   I(a,b) = 8\pi   \left(1+\frac{4}{3}\bar{s}  +\frac{215}{27} \bar{s}^2 \right) E_a E_b.     \label{Iab4}
\end{eqnarray}

It can be seen that {the $\lambda\varphi$ fluid} will increase the dark information of the Hawking radiation $I_0(a,b)=8\pi E_a E_b$, which is similar to the case of dark energy found in~\cite{MaChenSun:2018NPB}.
The change of the dark information due to {the $\lambda\varphi$ fluid} can be defined as
\begin{eqnarray}
   {\delta} I(a,b) &\equiv& I(a,b)- I_0(a,b) \nonumber \\
     &=&  \frac{\pi E_a E_b}{5M^3}   \, s
         +\frac{43 \pi  E_a E_b }{1920 M^6} s^2,
\end{eqnarray}
which increases with the scalar parameter $s$. With the evaporation of the black hole due to the Hawking radiation, the black hole mass decreases and so the additional dark information resulted from {the $\lambda\varphi$ fluid} increases for fixed $E_a$, $E_b$ and $s$. This is opposite to the case of dark energy~\cite{MaChenSun:2018NPB}, where the result is ${\delta} I(a,b)= 128 \pi M^2 \Lambda  E_a E_b$. When the black hole mass reduces to its minimum  $M_{\text{min}}(s)$ given by (\ref{Condition_M}), the additional dark information reaches its maximum:
\begin{eqnarray}
   {\delta} I(a,b)
     &=&   \frac{9^3  \Big[43\times 3^5-2^7 \times\Gamma^3 \left(-\frac{1}{3}\right)\Big] \pi }
              {5\times2^7 \times\Gamma^6\left(-\frac{1}{3}\right)} E_a E_b \nonumber  \\
   &\simeq& 15.2 E_a E_b.
\end{eqnarray}
Note that this is just an approximate estimate since it is not a small quantity anymore compared with $I_0(a,b)\simeq 25.1 E_a E_b$.

\section{Conclusions and Discussions}\label{conclusion}

In this paper, we have investigated the spacetime structure, Hawking radiation spectrum, and dark information of a spherically symmetric black hole in the background of {the $\lambda\varphi$ fluid.}

Firstly, with {the Langrangian of the $\lambda\varphi$ fluid (\ref{Lagrangian2})}, we obtained a spherically symmetric hair black hole solution with a positive scalar parameter $s$ and mass $M$. This solution has the same asymptotic behavior at spatial infinity as the Schwarzschild black hole except for a high-order correction from {the $\lambda\varphi$ fluid}, but has different behavior near the origin. However, the structure of its Penrose--Carter diagram is similar to the Schwarzschild case. For a fixed scalar parameter $s$, the black hole mass has a minimum (\ref{Condition_M}). The total energy of the black hole is $M$. The temperature and entropy of the black hole entropy were calculated. It was found that the entropy is not equal to one quarter of the area of the horizon.

Secondly, we used the statistical mechanical method based on canonical typicality~\cite{MaChenSun:2018EPL,MaChenSun:2018NPB} to calculate the nonthermal black hole radiation spectrum. This spectrum describes the statistical distribution of particle's energy after the particles cross the horizon through the Hawking process. The analytic result was given by (\ref{radiationSpectrum})--(\ref{chi2}) for general and small scalar parameter $s$. The exact numerical result was shown in Figure~\ref{Figprhs}. It was found that, during the Hawking radiation, the distribution probability of the radiation will increase slowly at a very long stage, and then the radiation will be accelerated rapidly at the late stage of the radiation. At last, the black hole mass is reduced to its minimum, but the radius is reduced to zero. The behavior of the radiation at the last stage is very different from the Schwarzschild black hole. It was also shown that {the $\lambda\varphi$ fluid} will speed up the Hawking radiation process and so reduce the lifetime of the black hole.

Lastly, we calculated the dark information reduced by {the $\lambda\varphi$ fluid by considering the information correlation between the radiated particles.} Such information correlation originates from the nonthermal radiation of the black hole and can not be measured locally. The result shows that, similar to the case of dark energy~\cite{MaChenSun:2018NPB}, the dark information of the Hawking radiation is also increased by {the $\lambda\varphi$ fluid}.
However, opposite to the case of dark energy, the dark information added by {the $\lambda\varphi$ fluid} increases during the evaporation of the black hole.

Note that, in our analyses, we did not consider the change of the scalar parameter $s$. If we consider the reduction of the parameter $s$ and the mass $M$ simultaneously, then we know from Figure~\ref{rhMSrelation} and Equation (\ref{M_rh}) that the horizon radius, the mass, and the scalar parameter can vanish at the last stage of the black hole. However, for a fixed scalar parameter, when the radius of the black hole shrinks to zero, the mass does not. This can also be seen from Figure~\ref{rhMSrelation}. This is very different from the case of the Schwarzschild black hole. According to Equation (\ref{entropySb}), the remnant state of the black hole carries non-vanishing entropy. According to Equation~(\ref{Rabcd2}) and Figure~\ref{FigRr}, it can be seen that the remnant state without a horizon is in fact a naked singularity. If the weak cosmic censorship conjecture is true, then the naked singularity should be hidden behind the event horizon, and thus the horizon radius should be stopped to shrink to zero by {some mechanisms} such as quantum gravity.

\section*{Acknowledgements}
Y.-X.L. would like to thank Hong L$\ddot{\text{u}}$, Hai-Shan Liu, Hai-Tang Yang, and Peng Wang for helpful discussions. The authors thank the anonymous referees for their valuable suggestions and comments.
This work was supported by the National Natural Science Foundation of China (Grants Nos. U1930403, 11875151, 12075103, and 12047501).


\end{document}